\documentclass{article}

\PassOptionsToPackage{numbers, compress}{natbib}

\usepackage[final]{ewrl_2024}

\usepackage[utf8]{inputenc} 
\usepackage[T1]{fontenc}    
\usepackage{hyperref}       
\usepackage{url}            
\usepackage{booktabs}       
\usepackage{amsfonts}       
\usepackage{nicefrac}       
\usepackage{microtype}      
\usepackage{xcolor}         

\usepackage{graphicx}
\usepackage{appendix}
\usepackage{subcaption}
\usepackage{amsmath}
\usepackage{wrapfig}
\usepackage{comment}

\title{Generalisation to unseen topologies:\\Towards control of biological neural network activity}

\author{
    Laurens Engwegen$^{1,2*}$\quad
    Daan Brinks$^{2*}$\quad
    Wendelin B\"ohmer$^{1}$
    \\
    $^*$Corresponding authors, 
    $^1$Department of Intelligent Systems,
    $^2$Department of Imaging Physics,
    \\
    Delft University of Technology,
    Delft, 2628XE, The Netherlands
    \\
    \texttt{\{l.r.engwegen,d.brinks,j.w.bohmer\}@tudelft.nl}
}

\begin{document}

\maketitle

\begin{abstract}

Novel imaging and neurostimulation techniques open doors for advancements in closed-loop control of activity in biological neural networks. This would allow for applications in the investigation of activity propagation, and for diagnosis and treatment of pathological behaviour. Due to the partially observable characteristics of activity propagation, through networks in which edges can not be observed, and the dynamic nature of neuronal systems, there is a need for adaptive, generalisable control. In this paper, we introduce an environment that procedurally generates neuronal networks with different topologies to investigate this generalisation problem. Additionally, an existing transformer-based architecture is adjusted to evaluate the generalisation performance of a deep RL agent in the presented partially observable environment. The agent demonstrates the capability to generalise control from a limited number of training networks to unseen test networks.
\end{abstract}

\section{Introduction}

Training reinforcement learning (RL) controllers for real-world applications is challenging. 
On the one hand, the amount of training interactions required by modern deep RL methods prohibits learning in the wild. 
On the other hand, simulated environments always differ in one way or another from the real world, and the trained neural networks generalise to this difference in unexpected ways.
The gap between the training environment and the real world is particularly large in biotechnology, where the controller faces noise, simplified simulation models, incomplete information about many simulation parameters, variability between individual instances of the entities to be controlled, and limited lifetimes of those entities. 
Meaningful generalisation over all of these factors is a key challenge in achieving optimal control in biotechnological applications, which would allow leveraging reinforcement learning to tackle some of the most intractable societal problems including green energy storage; mediation of climate change effects; and diagnostics and therapeutics for neurological and oncogenic disorders.

In this paper, we investigate the topological generalisation of deep RL architectures for closed-loop control of activity in networks of biological neurons. 
We first introduce an environment in which neuronal activity is simulated and an agent is tasked with controlling the activity in networks of biological neurons. The challenge lies in the fact that (1) each network has a different topology of which the connections between neurons can not be observed, and (2) the agent must generalise control to network topologies that are not seen during training.
A transformer-based architecture \citep{sensoryneuron} shows the capability to infer connections between neurons based on observed activity, necessary for generalisable control, indicating that this architecture can be effective in domains where dynamics are governed by non-observable graph-like structures. 
The experiments show that an RL agent with this architecture is successful in the multitask setting, where baseline architectures fail to obtain high performance on the test set.
The environment that is being introduced allows for the investigation of more factors of variation that play a role in real tissue, such as different levels of noise, a varying number of (non-)observable neurons, and differences in the dynamics that govern activity propagation.
While we use these biological systems as an example, the developed concepts hold interest for control in other environments with incomplete information, heterogeneity and limitations on data and training availability.

This paper takes a first step towards the application of RL to investigate neuronal network dynamics on a circuit level through generalisable control. The main contributions are the following:
\begin{itemize}
    \item We introduce an environment in which networks of neurons are procedurally generated, and simulated with a biophysical model for action potential initiation and propagation.
    \item We adjust an existing transformer-based architecture to control neuronal activity in different neuronal networks.
    \item We show empirically that this architecture can generalise control to neuronal networks with unseen topologies, outperforming renowned methods for environments in which the state of the system is partially observable.
\end{itemize}

\section{Reinforcement Learning and Generalisation}

A Markov Decision Process (MDP) \citep{suttonbarto} is often used to model decision-making problems. In many real-world applications, the control agent can not fully observe the state of the environment. The MDP framework can be extended to a Partially Observable Markov Decision Process (POMDP) \citep{pomdp} that is defined by a tuple $\mathcal{M} = \langle \mathcal{S}, \mathcal{A}, \Omega, \mathcal{T}, \mathcal{R} ,\mathcal{O}, \rho_0 \rangle$, consisting of a set of states $\mathcal{S}$, a set of actions $\mathcal{A}$, and a set of observations $\Omega$. Additionally, it contains a transition function, $\mathcal{T}(s'|s,a)$, that maps (current) state $s \in \mathcal{S}$ and action $a \in \mathcal{A}$ to a probability distribution over (next) states $s' \in \mathcal{S}$, a reward function $\mathcal{R}: \mathcal{S}\times\mathcal{A}\times\mathcal{S} \rightarrow \mathbb{R}$, and an observation function that maps a state to a probability distribution over observations $\mathcal{O}: \mathcal{S}\times\Omega\rightarrow [0,1]$. In any state, the agent can only observe the observation that is produced from that state by $\mathcal{O}$. The last element of the tuple, $\rho_0$, is the initial state distribution. 

The goal of RL is to learn a policy $\pi$, with which an agent chooses actions based on available information,
that maximises the expected (discounted) cumulative reward $J = \mathbb{E}_\pi[ \sum_{t=0}^{\infty} \gamma^t r_t ]$, with discount factor $\gamma \in [0,1)$ and the reward at time step $t$ in the environment $r_t = \mathcal{R}(s_t, a_t, s'_t)$. A POMDP can be solved as an induced belief-state MDP in which the complete action-observation history (AOH), which is defined at time step $t$ as $\tau_t = (o_0, a_0, \dots, o_{t-1}, a_{t-1}, o_t)$, is summarised by the agent \citep{pomdp} to be able to construct a belief over the current state it is in. In this way, a policy $\pi(a|\tau)$, mapping $\tau$ to a probability distribution over actions, can be learned with any RL algorithm. 

Q-learning \citep{qlearning, doubleqlearning} is often used to learn the optimal Q-function that maps an action $a$ and $\tau$ to the expected cumulative reward when following policy $\pi$: $Q^{\pi}(\tau, a) = \mathbb{E}_{\pi}[ \sum_{k=0}^{\infty} \gamma^t r_{t+k} | \tau_t = \tau, a_t = a]$. If we find the optimal Q-function, $Q^*(\tau, a) = \mathrm{max}_{\pi}Q^{\pi}(\tau,a)$, an optimal policy $\pi^*$ can be derived by greedily taking actions with respect to these Q-values. 
When dealing with large state and/or action spaces, it is infeasible to learn Q-values for each AOH and action pair individually. In this case, we can learn a parameterised Q-function, e.g. a Deep Q-Network (DQN) \citep{DQN} in MDPs, and a Deep Recurrent Q-Network (DRQN) \citep{DRQN} in POMDPs.  In the latter, a recurrent neural network (RNN) is exploited to learn a latent representation of $\tau$.

In real-world environments an agent will often encounter many variations (called \textit{contexts}) of a (PO)MDP, and training an agent on each of these variations individually is usually infeasible. In such cases, an agent is required to learn a policy that solves all training contexts, in the hope that it also generalises control to new contexts that are not seen during training. Such contexts determine the task of an RL agent, e.g. the configuration of objects in the environment or the level of a game. The Contextual Markov Decision Process (CMDP) \citep{CMDP} framework explicitly allows to evaluate generalisation capabilities \citep{kirk}. In a CMDP, the state space is factored into the \textit{underlying} state space and the context space $\mathcal{S}: \mathcal{S}' \times \mathcal{C}$. When the context is partially observable, the CMDP can be thought of as a POMDP in which the context is fixed throughout an episode. The CMDP allows us to construct a set of training and testing POMDPs, $\mathcal{M}|_{\mathcal{C}_{\mathrm{train}}}$ and $\mathcal{M}|_{\mathcal{C}_{\mathrm{test}}}$ respectively, where $\mathcal{C}_{\mathrm{train}}, \mathcal{C}_{\mathrm{test}} \subseteq \mathcal{C}$ and $\mathcal{C}_{\mathrm{train}} \cap \mathcal{C}_{\mathrm{test}} = \emptyset$. 
To this extent, after training the (zero-shot) generalisation capabilities of an agent can be evaluated on an unseen test set of POMDPs.

\section{Neuronal activity}

\begin{wrapfigure}{r}{0.5\textwidth}
  \vspace{-6mm}
  \begin{center}
    \includegraphics[width=0.48\textwidth]{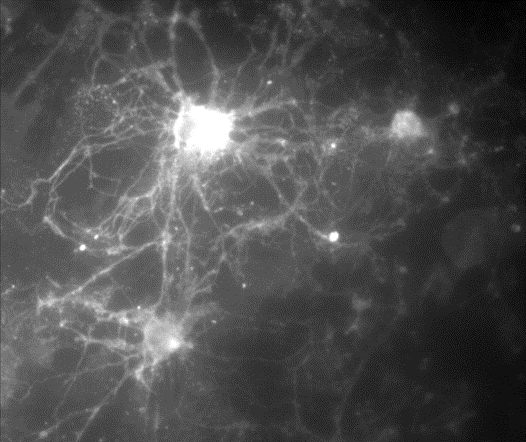}
  \end{center}
  \caption{Voltage image of neurons grown \textit{in vitro}. When the field of view and number of neurons increases, it quickly becomes infeasible to study neuronal dynamics in detail.}
  \label{fig:voltage_image}
\end{wrapfigure}

Biological neurons communicate with each other through action potentials. Disturbed generation and propagation of these signals is associated with various neurological disorders, such as Parkinson’s disease, epilepsy, and anxiety disorders, among others \citep{optogenetic_applications,epilepsy_neuralcircuit}. Diagnosis can be difficult and time-consuming, and treatment is even more complex. 
Established recording modalities, such as fMRI, EEG and microelectrode arrays, are widely used to study brain tissue, both \textit{in vitro} or \textit{in vivo}. These modalities are all subject to a relatively low spatial or temporal resolution, which deprives researchers from studying brain tissue on a cellular or circuit level. Advances in voltage imaging \citep{voltimg_first, voltimg_recentreview} have shown potential to overcome these hurdles, enabling researchers to study action potential propagation on a network or cellular level with high resolution.
Optogenetics \citep{optogenetics1, optogenetics2, optogenetics3} subsequently allows for the manipulation of activity in individual neurons with action potential precision.
These techniques provide promising avenues for machine learning to contribute to understanding, diagnosis and treatment through measuring, analysis and stimulation of neuronal electrical signalling. 
The high-dimensional and complex correlatory nature of neuronal signalling makes it a natural fit for machine learning to approach full understanding and control of neural network activity (Figure \ref{fig:voltage_image}).

RL agents, specifically, have the potential to manipulate neuronal activity to effectively unravel properties of the dynamical system or to induce specific activity patterns.
However, the applicability of RL is particularly limited by the fact that we cannot observe the complete state of the neuronal network, and by the limited lifetime of the sample inhibiting exhaustive training routines. 
Therefore, a key feature of RL agents as applied to neuronal activity must be the capability to transfer learned control from a set of training samples to new, unseen samples. 
As such, when we want to control the activity in networks of biological neurons, we are concerned with a generalisation problem: every network of neurons has a different topology, we want to induce different activity patterns for different neurons in each network, and we cannot (re)train an agent on every new network that is to be investigated. This problem can be formulated as a CMDP, where the context is partially observable as the connections between neurons in a network can not be observed by the agent. The desired activity pattern and the \textit{target} neuron(s) can be defined by an expert and therefore constitute the observable context.
By constructing training and test sets of such partially observable contexts, the generalisation performance of DRQN variations with different architectures can be investigated.

\section{Related Work}

\subsection{Neuronal simulation}

To study neuronal network dynamics \textit{in silico}, various models for the initiation and propagation of activity have been developed. Oftentimes, neuronal dynamics in individual neurons are defined by an integration process of incoming activity, together with a mechanism that initiates an action potential when the neuron reaches a critical threshold voltage \citep{neuronaldynamics_book}. The Leaky Integrate-and-Fire (LIF) model \citep{LIF_introduction}, and its various extensions, is widely used to model spiking neurons \citep{biophysicsofcomp, brunel2000dynamics, izhikevich_extension, adapexp_LIF}, due to its computational efficiency and the emergent realistic behaviour in networks of LIF neurons (with the right hyperparameter values). 

Self-organised criticality (SOC) is a concept that is used to describe and understand some of the scaling dynamics in complex natural systems \citep{original_criticality, soc_forestfire}.
\citet{braincriticality_original} showed for the first time that activity propagation in cortical neurons can be described by equations that govern avalanches, indicative of critical behaviour where system dynamics are poised between order and randomness. Since then, measures for critical behaviour and phenomena through which critical behaviour is established, such as plasticity in synapses, have been investigated \citep{SOC_theory, SOC_synplast}. Various \textit{in silico} neuronal models combined LIF neurons with short-term synaptic depression to induce SOC in simulations \citep{SOC_LIFSTSD_1, SOC_LIFSTSD_2, SOC_reproduced}. In the current work, we implement a simplified version of the model introduced by \citet{SOC_reproduced}. 
See Appendix \ref{app:simulation} for details of the implementation.

Lastly, network level dynamics are predominantly determined by the connectivity of the network. Various graph theoretical studies on neuroimaging data have demonstrated that different networks of functional and structural connectivity have \textit{small-world} properties \citep{original_smallworld, smallworld_brainnetworks, smallworld_brainnetworks2}. While elaborate methods are being developed to generate realistic connectivity matrices on basis of little experimental data \citep{convolutive_model}, we prioritise a general model for the generation of connectivity matrices that does not overfit to a specific type of tissue. Hence, small-world networks will be generated using simple rules, similar to \citep{SOC_reproduced}.

\subsection{Closed-loop control of neuronal activity}

Control of neuronal activity has shown to be useful for therapeutic and research purposes. Deep brain stimulation (DBS) \citep{DBS} is a neurosurgical method that is currently being used for the treatment of Parkinson's and Alzheimer's disease, and the prevention of epileptic seizures, among others. DBS is typically applied in an open-loop fashion, where fixed patterns of stimulation are induced to reduce pathological symptoms. Attempts have been made to close the loop to reduce adverse effects and increase the efficiency of DBS systems. Moreover, the dynamical nature of neuronal systems often requires an adaptive strategy to be effective in real-life systems. \citet{closedloop_sci} developed a closed-loop neuromodulation approach to improve recovery from spinal cord injury. This was possible by utilising synaptic plasticity and, thus, required adaptive control. Other studies presented RL approaches to minimise stimulation energy needed to control epileptic seizures \citep{RL_seizurecontrol}, and treat Parkinson's disease \citep{RL_parkinsons}, in computational models of those disorders. Both studies trained and tested agents on the same network, and did not investigate generalisation to different instances of diseased networks.

DBS traditionally utilises electrical stimulation and recordings through injected electrodes, which can cause injury and hamper real-time readouts of activity. Advancements in adaptive, closed-loop, DBS are motivated by reduced risk of such effects, and accelerated by cutting-edge technologies like voltage imaging and optogenetics that allow for real-time feedback and control. Closed-loop optogenetic control systems have already been developed. \citet{closedloop_alloptical}, for example, conditioned \textit{target} neurons on \textit{trigger} neurons by means of photostimulation and demonstrated long-lasting changes in activity of neuronal ensembles. This study suggested the possibility to correct aberrant activity patterns in diseased conditions by reconfiguration of functional connectivity. Other studies demonstrated closed-loop optogenetic control to suppress pathological seizure-like patterns \citep{closedloop_nonhumanprimates}, or to restore bladder function \citep{closedloop_alloptical_bladderrat}.

Work on closed-loop control of \textit{in silico} neuronal systems that is most closely related to our paper was presented by \citet{close_work}, in which stochastic LIF neurons were simulated and spiking patterns had to be produced by deep RL agents. They (re)trained and tested agents on individual networks of neurons and made the strong assumption that external stimulation (in their case both excitatory and inhibitory) could override internal activity propagation in the network. Furthermore, they learned a separate policy for each neuron in the network, which limits the scalability of their approach. In the current work, we do not make their assumption, and a single policy is learned to control activity in neuronal networks. Besides, we investigate the performance of our RL agent on \textit{any} network, rather than on a single instance. The latter is important for practical applications, where neuromodulation approaches to investigate or reinforce neuronal network activity are desired to be effective immediately. This requires transferring knowledge and, thereby, generalising control to unseen networks.

\subsection{Multitask learning}
The core objective of multitask RL is to improve generalisation performance \citep{MTRL_survey}. When training on a distribution of tasks, agents need to leverage domain-specific, rather than instance-specific information to learn a general representation of the system. This representation should be informative and suitable for any task the agent can find itself in. To be consistent, we will refer to tasks as contexts, as introduced in the previous section. \\
As described by \citet{kirk}, the generalisation performance (or generalisation gap) can be defined by the difference between an agent's performance on a training set and a test set of contexts. A popular benchmark for the development and evaluation of multi-tasking agents is ProcGen \citep{procgen}, that consists of different environments in which different contexts are procedurally generated. In the current work, an environment in which contexts are procedurally generated is introduced to be able to evaluate an agent's generalisation performance, data efficiency, scalability, and robustness, with respect to different contexts.

\subsection{Architecture}

To generalise control to different, unseen topologies in which the edges between nodes can not be observed, an architecture through which the agent can infer relationships between nodes is required. \citet{sensoryneuron} introduced a combination of RNNs and a transformer \citep{vaswani2017attention} to construct a permutation invariant agent. They showed that their architecture is invariant to permutations of the input features during an episode. We adopted and adjusted this architecture, using the self-attention mechanism, to induce permutation equivariance in the RL agent. Transformers are starting to be adopted in RL, specifically by viewing the objective as a sequence modelling problem \citep{decisiontransformer, offlineTransformer}, or to learn a world model of the environment \citep{transdreamer}. Here, we use it as a representation learning model to capture dependencies between specific objects (neurons in a graph). The self-attention mechanism is at the heart of the transformer, through which attention scores are calculated as $A = \sigma (QK^T/\sqrt{d}) V$, where $Q,K,V \in \mathbb{R}^{n \times d}$, are the query, key and value matrices that are projections of the input $X \in \mathbb{R}^{n \times d}$, such that $Q = XW_Q$, $K = XW_K$ and $V = XW_V$, with learnable parameter matrices $W_Q, W_K$ and $W_V$, and $\sigma(\cdot)$ a row-wise softmax function.\\
Where \citet{sensoryneuron} showed robustness to permutations of the input on a single task, we experiment with the adjusted version of their architecture in a multitask setting. Moreover, we evaluate generalisation to different topologies, rather than permutations of the same topology.
The use of transformers has shown to be effective in other domains, for example in agent-agnostic control \citep{amorpheus}, where the self-attention mechanism could be exploited to control robots in the absence of morphological information, but where generalisation to new morphologies was limited.

\section{Method}

The environment and conceivably effective architectures are described here. More details can be found in Appendix \ref{app:simulation}, \ref{app:context_generation} and \ref{app:context_encoding}

\subsection{Environment}\label{subsec:environment}

The goal in this environment is to create a specific target spiking pattern in a target neuron, by manipulating the activity in other neurons. Performance is evaluated in a multitask setting. The agent has to create arbitrary spiking patterns in neuronal networks with different topologies of which the connections are can not be observed: relevant relationships between neurons have to be inferred through interaction. The elements of the partially observable CMDP are briefly explained.

\paragraph{States and observations.} States in the environment are defined by a multitude of variables, such as the activity of the different neurons, their underlying topology, and the simulation parameters of activity propagation. Only the activity of neurons (membrane potential at the soma), however, can be observed. In the current work, we omit rendering images. It is assumed that we can accurately transform a voltage image to activity measures (i.e. the membrane potential) for each neuron by making use of available segmentation and analysis software. In a simulated environment with $n$ observable neurons, an observation $o \in \mathbb{R}^n$ thus contains the membrane potential of the neurons.

\paragraph{Actions.} The agent can activate (i.e. initiate an action potential in) any of the $n$ observable neurons or do nothing (indicated by $-1$): $\mathcal{A} = \{-1,0, ..., n-1\}$. In the current set-up, the neuron that the agent needs to control is excluded from the set of possible actions in order to make the task more challenging in smaller networks (e.g. $n=8$)\footnote{In larger, complex networks, however, activating the target neuron could have a significant effect on its future activity, and could thus be allowed while ensuring challenging tasks.}.

\paragraph{Transitions.} The environment simulates LIF neurons with short-term synaptic depression \citep{SOC_reproduced, SOC_LIFSTSD_1, SOC_LIFSTSD_2}. Details on the simulation procedure can be found in Appendix~\ref{app:simulation}. In short, the neuronal networks consist of excitatory and inhibitory neurons (with a ratio of 4:1 on average). Active excitatory neurons will propagate their activity to connected neurons, delivering a positive input current in the next time step. Active inhibitory neurons will propagate a negative input current. When an agent chooses to activate a neuron, this neuron will receive a strong positive (external) input current. In the present setup, transitions are deterministic and only depend on the current state and action.

\paragraph{Procedural context generation.} Contexts in the environment consist of (1) $c_{\mathrm{topology}}$, the \textit{non-observable topology} of the biological neuronal network, (2) $c_{\mathrm{neuron}}$, the \textit{observable target neuron} that the agent has to control, and (3) $c_{\mathrm{pattern}}$, the \textit{observable target spiking pattern} that the agent has to create in the target neuron. The target neuron is defined by the ID of the neuron: $c_{\mathrm{neuron}} \in \{0, ..., n-1\}$. The target pattern is a binary vector that indicates at which time steps the target neuron must be spiking: $c_{\mathrm{pattern}} \in \{0,1\}^N$ during an episode of $N$ time steps. The procedural generation of the elements that constitute the partially observable context, $c = [c_{\mathrm{topology}}, c_{\mathrm{neuron}}, c_{\mathrm{pattern}}]$, is explained in detail in Appendix~\ref{app:context_generation}. To test generalisation to different (non-observable) topologies, the training and test context sets consist of unique topologies only:
$\forall c, c' \in \mathcal{C}_{\mathrm{train}} \cup \mathcal{C}_{\mathrm{test}}, (c \neq c') \rightarrow (c_{\mathrm{topology}} \neq c'_{\mathrm{topology}})$.

\paragraph{Rewards.} A reward only depends on the current state $s_t$. If the activity (i.e. spiking or not spiking) of the target neuron $c_{\mathrm{neuron}}$ corresponds to the desired spiking pattern at the current time step, $(c_{\mathrm{pattern}})_t$, the agent receives a positive reward. To reward correct spikes as much as correct inactive time steps, the active reward $R_A$ is scaled by the number of spikes $m$ in the target pattern: $R_A = \frac{R_{\mathrm{max}}}{m}$, and the inactive reward $R_I$ by the number of inactive time steps $N-m$: $R_I = \frac{R_{\mathrm{max}}}{N-m}$. When the target neuron is either incorrectly active or inactive, the agent receives a reward of 0.

For the experiments discussed in this work, various (possible) properties of realistic voltage images are not considered, such as non-observable neurons, neurons with different dynamics, background noise, photobleaching effects, and stochastic activity propagation, among many others. Nonetheless, the environment can be (and has been) extended to include such properties, each of which comes with its own challenges for the RL agent. In this paper, however, we specifically investigate generalisation to different topologies and, therefore, reduce the environment's complexity.

\subsection{Agent architectures}

\begin{figure}[t!]
    \centering
    \includegraphics[width=\textwidth]{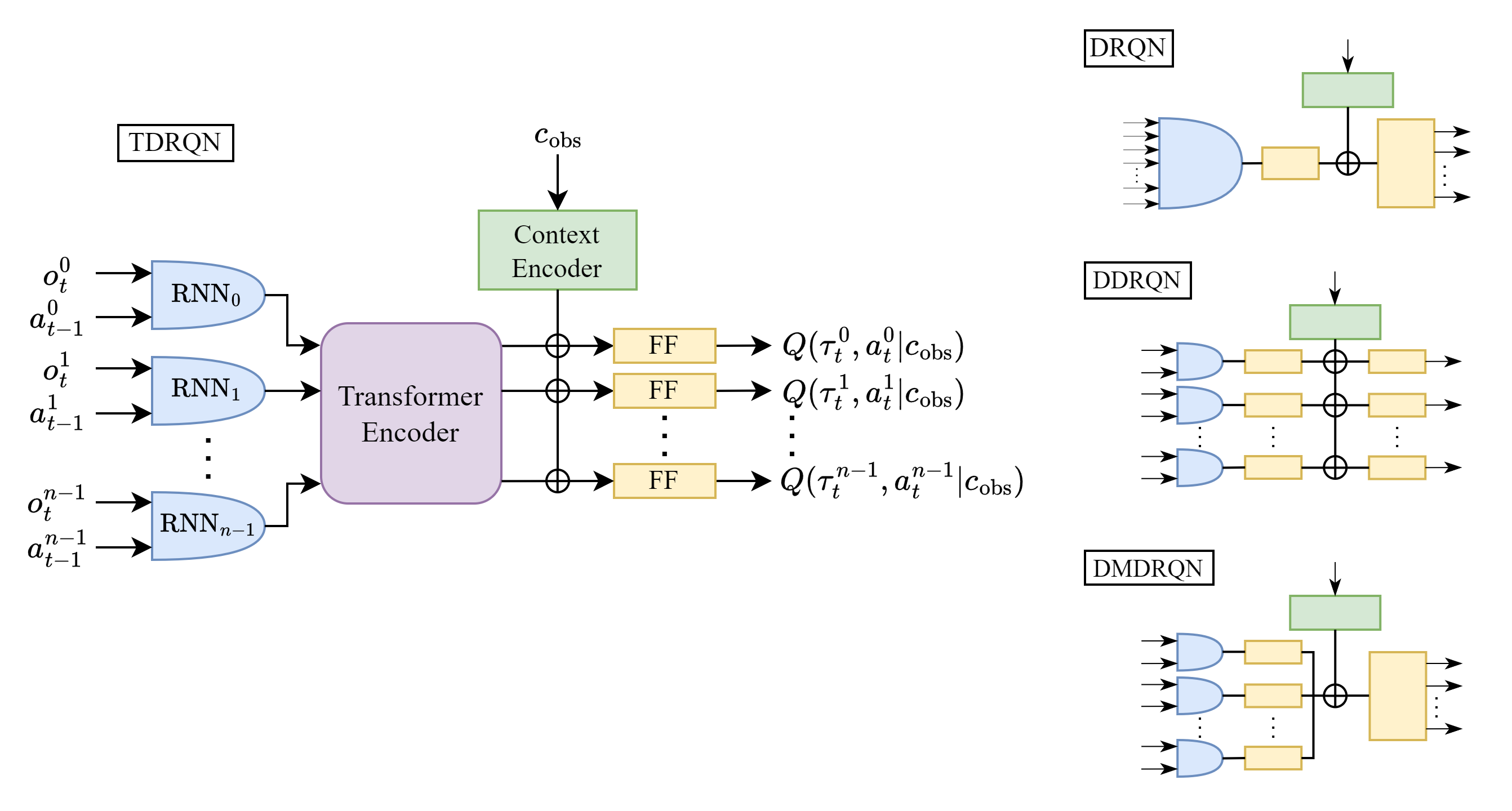}
    \caption{The architectures with which we experiment. Different RNNs (in TDRQN, DDRQN and DMDRQN) share the same parameters, but keep track of different hidden states. FF blocks placed below each other correspond to layers with the same parameters, but applied to different inputs. The concatenation operation is indicated with $\oplus$. The components of the (baseline) architectures that are schematically visualised on the right correspond to those in TDRQN with the same colour.}
    \label{fig:architecture_diagram}
\end{figure}

The major challenge in the introduced environment is the combination of generalisation to unseen contexts and the partially observable nature of the CMDP.
To this extent, we investigate what inductive biases in the agent's architecture result in good generalisation performance.
Recurrent neural networks are often used to learn a representation of the AOH, $\tau$, in POMDPs, as in DRQN \citep{DRQN}.
Different recurrent architectures for the parameterisation of Q-functions are implemented in this work to evaluate their generalisation performance on the introduced partially observable CMDP. Those architectures are schematically visualised in Figure \ref{fig:architecture_diagram}. We use Long Short-Term Memory (LSTM) as the recurrent component in each of these architectures.
The target neuron and target pattern are the part of the context that is observable, to which we refer as $c_{\mathrm{obs}}$.\\
The baseline \textbf{DRQN} encodes the AOH of the system with a single RNN.
The observable context is encoded by feed-forward (FF) networks (see Appendix \ref{app:context_encoding} for more details) and concatenated to the latent representation of $\tau$. This embedding is mapped to Q-values for each action.

\citet{sensoryneuron} distribute the mapping of the AOH over different RNNs. In the current environment, this can be achieved by mapping the observation and previous action, $o_{t}^{i}$ and $a_{t-1}^{i}$, at time step $t$ of each individual neuron $i$ with its own LSTM. 
In practice, we use one LSTM (i.e. one set of trainable parameters), but keep track of a different hidden state for each neuron, which encodes the local AOH $\tau^i_t$ for that neuron.
The previous action, $a_{t-1}^{i}$, here indicates whether or not neuron $i$ was activated in the previous time step. 
A transformer encoder with self-attention is utilised to attend the encoding of local AOHs to each other, producing a latent embedding for each neuron that is equivariant to permutations of the neurons.
Consecutively, the context embedding is concatenated to those latent codes and mapped to Q-values: $Q(\tau^i_t, a^i_t|c_{\mathrm{obs}})$. We will refer to this architecture as \textbf{Transformer-based Deep Recurrent Q-Network (TDRQN)}.

To investigate the effect of the self-attention mechanism in the transformer block of the TDRQN on generalisation, we compare it against two other architectures. \textbf{Distributed Deep Recurrent Q-Network (DDRQN)} encodes local AOHs as in TDRQN, but applies feedforward layers to those embeddings (that are again combined with context embeddings). In this way, local AOHs are mapped to Q-values in a permutation equivariant manner. With this architecture, we can evaluate whether self-attention is crucial for (generalisable) control to learn to associate local AOH embeddings, or that we can effectively map the embeddings directly to Q-values with FF layers.

The last architecture for which we will report results, distributes the encoding of the AOH, again, over local LSTMs, but later combines (i.e. concatenates) these embeddings to allow for inference of relationships between local AOHs. This architecture will be called \textbf{Distribute-and-Mix DRQN (DMDRQN)}. Concatenating the local AOHs precludes permutation equivariance, but does allow the agent to directly associate different local AOH embeddings.

Although the AOH is distributed over different LSTMs and therefore permutation equivariant, it is still not straightforward for the agent to perform well in permuted tasks. At the start of an episode, namely, each local AOH is initialised in the same way and, thus, indistinguishable from the others. Only after interaction with the environment and observing activity in neurons, the agent can induce different AOHs from which relationships between neurons can be inferred. As a consequence, the agent can only establish the correct activity pattern at the very beginning of an episode by chance. As the reward function was designed in such a way that the agent receives a positive reward for correct activity at each time step, the maximum return in an episode can, thus, only be obtained in training contexts to which the agent would overfit.

\section{Results}

The introduced architectures are trained and tested on the two different sets of partially observable contexts, $\mathcal{C}_{\mathrm{train}}$ and $\mathcal{C}_{\mathrm{test}}$, to evaluate generalisation performance. Each set consists of 100 contexts. For the current experiments, networks contain $n=8$ neurons and connections can not be observed.\\
More detailed experimental information can be found in Appendix \ref{app:exp_setup}. Since the active and inactive rewards are weighted, as described in the previous section, the return of an episode in which an agent did not do anything is equal to 0.5 (which is why lower values are not shown in the Figures). 

\begin{figure}[htbp] 
  \centering
  \includegraphics[width=\linewidth]{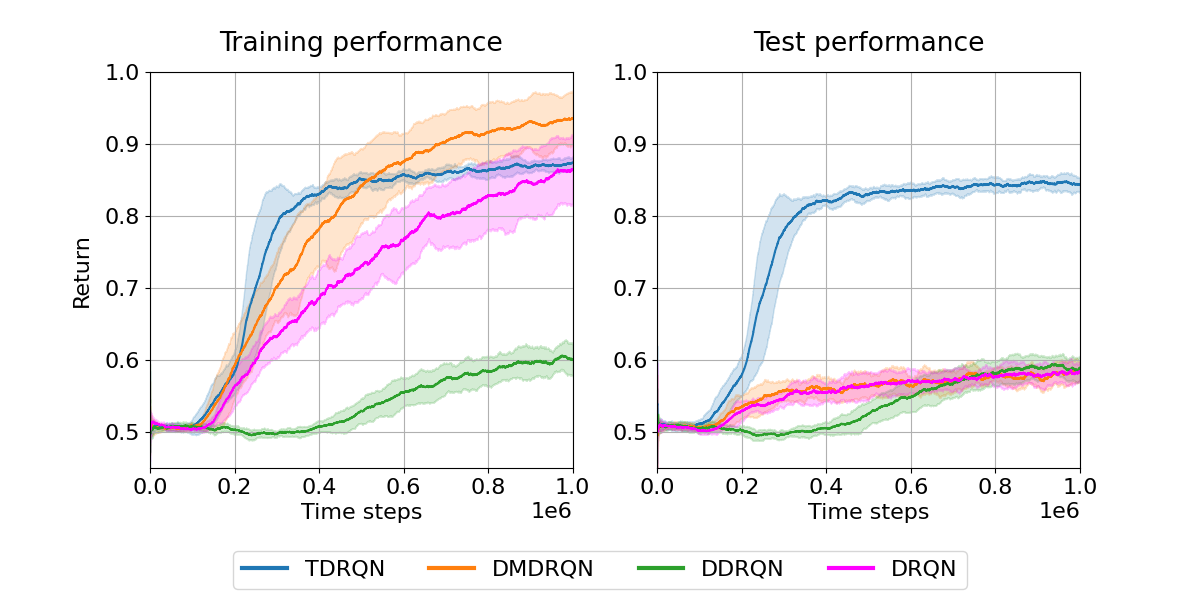}
\caption{Performance on 100 training (left) and 100 unseen test (right) contexts for TDRQN, DDRQN, MDDRQN, and DRQN. Performance is averaged over 10 seeds and the standard deviation is shown in the shaded area.}
\label{fig:compare_performance}
\vspace{-6mm}
\end{figure}

Figure \ref{fig:compare_performance} shows that TDRQN successfully generalises control to the test set, without any information about the topology of the networks that are to be controlled. 
The baseline DRQN, that encodes the AOH of the system as a whole, overfits to the training set. We speculate that this way of encoding the AOH does not allow the agent to form proper belief states outside of the training set: none of the states encountered during testing are encountered during training, which could complicate encoding proper belief states.
The states of individual neurons and the possible effects of neurons among themselves, however, do transfer from training to test set, despite varying global dynamics. By combining or mixing local AOHs, the agent could successfully infer the effect of (acting on) each of the neurons with respect to the task at hand, as shown with the results of the TDRQN agent.

\begin{wrapfigure}{r}{0.5\textwidth}
    \centering
    \includegraphics[width=0.48\textwidth]{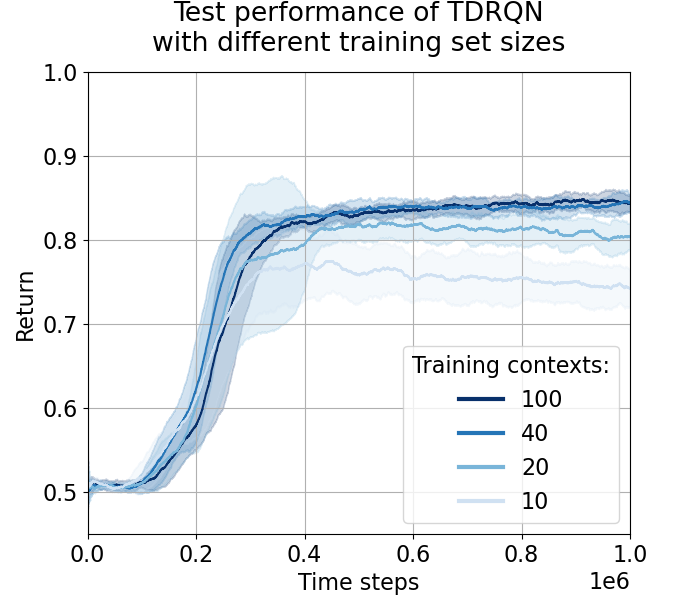}
    \caption{Performance on unseen test contexts of TDRQN with different amounts of training contexts. Performance is averaged over 10 seeds and the standard deviation is shown in the shaded area.}
    \label{fig:traintasks}
\end{wrapfigure}

The DMDRQN, in which local AOHs are concatenated together with the encoded observable context, does also overfit to the training data. This suggests that a permutation equivariant mixing procedure (as in TDRQN) is crucial for generalisation performance. Finally, the permutation equivariant DDRQN architecture shows poor training and testing performance. This shows that a mechanism to combine or mix the AOH of all neurons is required for the agent to infer the effects that different neurons have on each other.
These results indicate that the distribution of local action-observation histories in combination with an elaborate mechanism (self-attention) to infer relationships between activity patterns from local AOH embeddings is essential for generalisation to unseen topologies.

We additionally investigated the sample efficiency of the TDRQN agent, as this is an important indication for the applicability of RL agents in real-world environments. Specifically, when we want to learn to control activity in biological neuronal networks, it is time-consuming and costly to generate a sample to train on. Figure \ref{fig:traintasks} shows the test performance of TDRQN trained on training sets with different sizes. These results show that the agent already reaches high generalisation performance with only 40 training contexts, which would be in the feasible regime for real-world applications.

\section{Discussion}

In this paper, an environment that simulates biological neuronal networks on a circuit level was introduced. Although numerous abstractions have been made, neurons are simulated with an often used (LIF with synaptic plasticity) model for activity propagation based on biophysical properties. Besides, the environment allows to tackle some key challenges that arise in real-life biological tissue, such as its partial observability and the necessity to generalise control to unseen samples. To start addressing these issues, an existing transformer-based architecture was slightly adjusted and evaluated to control the activity in neuronal networks with procedurally generated contexts. Where it was already shown that such architecture can be robust to permutations of the input, here we have shown that it can generalise to unseen contexts while requiring a limited number of training samples. The transformer encoder effectively learns to associate the dependence between objects in an environment, in our case simulated neurons, from their local action-observation history.

The investigated architecture could be useful for generalisation in other domains with an underlying topological structure, such as traffic light control \citep{VanDerPol16, CabrejasEgea21}, or robotic control in a morphology-agnostic manner, i.e., to generalise control to different robots without the need of morphological information \citep{amorpheus}. Furthermore, it would be interesting to investigate its generalisation performance in the visual domain. 
For example, the experiments on visual tasks in the original paper of Tang and Ha \citep{sensoryneuron} could be extended to evaluate performance in different contexts, rather than in a permuted version of the same context.
However, the quadratic complexity of self-attention and the need to encode a separate local action-observation histories require a lot of compute and memory. It would, thus, be beneficial for scalability to develop more efficient architectures \citep[e.g.][]{Sun23}.

The presented experiments were performed on simulations that are simplifications of actual neuroimaging data that we can currently acquire. Although these experiments tackled the fundamental problem of partial observability of the context of the task, realistic features of the real-world setting could be introduced to complicate control. For example, voltage images are inherently noisy, spiking behaviour is naturally stochastic and governed by a multitude of synapses rather than a single connection, and we can often not observe all neurons in the sample. Future extensions of the environment could incorporate such aspects to challenge RL agents along the way to investigation of neuronal tissue through closed-loop control. Nonetheless, the introduced environment enables experimentation in different RL research areas, such as multi-agent RL, causality, and meta-learning, besides generalisation in partially observable environments, with potential applications in neuroscience.

\section*{Acknowledgements}
We thank Qiangrui Dong for generating the voltage image in Figure \ref{fig:voltage_image}, and He Tian and Adam Cohen for the gift of QuasAr6a.
This project was funded by the Delft AI initiative within the BIOlab. This work was also partially funded by the Dutch Research Council (NWO) project {\em Reliable Out-of-Distribution Generalization in Deep Reinforcement Learning} with project number OCENW.M.21.234. DB acknowledges support from an ERC Starting Grant (850818 - MULTI-VIsion) and the Convergence Health and Technology (Integrative Neuromedicine flagship).

\bibliographystyle{unsrtnat}
\bibliography{bibliography}

\newpage

\appendix

\section{Simulating neuronal networks}\label{app:simulation}

Neurons are simulated with the LIF model and short-term synaptic depression.
The current simulation method is a simplification of \citep{SOC_reproduced}. At each time step $t$ (of 1ms) the membrane potential $V_j$ of neuron $j$ is updated by the following differential equation:
\begin{equation}\label{eq:sim}
    \frac{dV_j}{dt} = \frac{V_j-V_r}{RC} + \frac{1}{C} \left( I^{j}_{e}(a,t) + \sum_i \sum_k H(p_r U_k(t) - 0.5) I_{in}^j(i, t) \right)
\end{equation}
where $V_r$ indicates the resting potential, $R$ the resistance, and $C$ the capacitance of the cell membrane. In the current setup, when an agent acts in the environment, it can induce external activity $I^{j}_{e}(a,t) = w_{e}\delta(j-a)$ of input current $w_e$, when its action $a$ corresponds to neuron $j$.
The input current from other neurons $i$ in the network to neuron $j$ is defined by:
\begin{equation}
    I_{in}^{j}(i, t) = w_{in} \delta (t-t_{s}^{i})G_{i,j}
\end{equation}
with  $w_{in}$ the weight (synaptic strength) of incoming internal input. $t_{s}^{i}$ contains the (last) spike time of neuron $i$, and the connectivity matrix $G$ indicates whether there is an excitatory ($1$), inhibitory ($-1$), or no ($0$) connection between two neurons.\\
$H$ is the heavy-side step function and $p_r$ is a probability that influences the possibility of a pre-synaptic synapse releasing its vesicles. Short-term synaptic depression is governed by $U_k$, which relates to the availability of vesicles and neurotransmitters at pre-synaptic site $k$:
\begin{equation}
    U_k(t) = 1 - e^{- \frac{t - t_{r}^{k}}{\tau_R}}
\end{equation}
where $t_{r}^{k}$ indicates the last time step at which vesicles were released at site $k$, and $\tau_R$ is the time constant for the uptake of vesicles and neurotransmitters.\\
Whenever a neuron's membrane potential reaches the threshold $V_{\mathrm{th}}$, it emits a spike or action potential. Directly after this spike, the membrane potential is reset to the resting potential and the neuron enters a refractory period, $\tau_{\mathrm{rp}}$, during which it is unable to spike (even from external input).

\begin{table}[htbp]
    \caption{Hyperparameters used in simulation of neuronal networks.}
    \label{tab:simulation_hyperparameters}
    \centering
    \begin{tabular}{l|r}
        \textbf{Hyperparameter} & \textbf{Value} \\
        \hline\hline
        Resting potential, $V_r$ & -70 mV \\
        Spike threshold, $V_{\mathrm{th}}$ & -50 mV \\
        Refractory period, $\tau_{\mathrm{rp}}$ & 1 ms \\
        Resistance, $R$ & $\frac{2}{3} \times 10^9\ \Omega$ \\
        Capacitance, $C$ & $3 \times 10^{-11}$ F \\
        External input strength, $w_e$ & $1200$ pA \\
        Internal synaptic strength, $w_i$ & $600$ pA \\
        Vesicle release probability, $p_r$ & 0.65 \\
        Uptake time constant, $\tau_R$ & 3 ms \\
    \end{tabular}
\end{table}

\section{Procedural context generation}\label{app:context_generation}
The procedural generation of contexts consists of three steps: network topology generation, target neuron assignment and target spiking pattern creation.

\begin{figure}[htbp]
\begin{subfigure}{.33\textwidth}
  \centering
  \includegraphics[width=.95\linewidth]{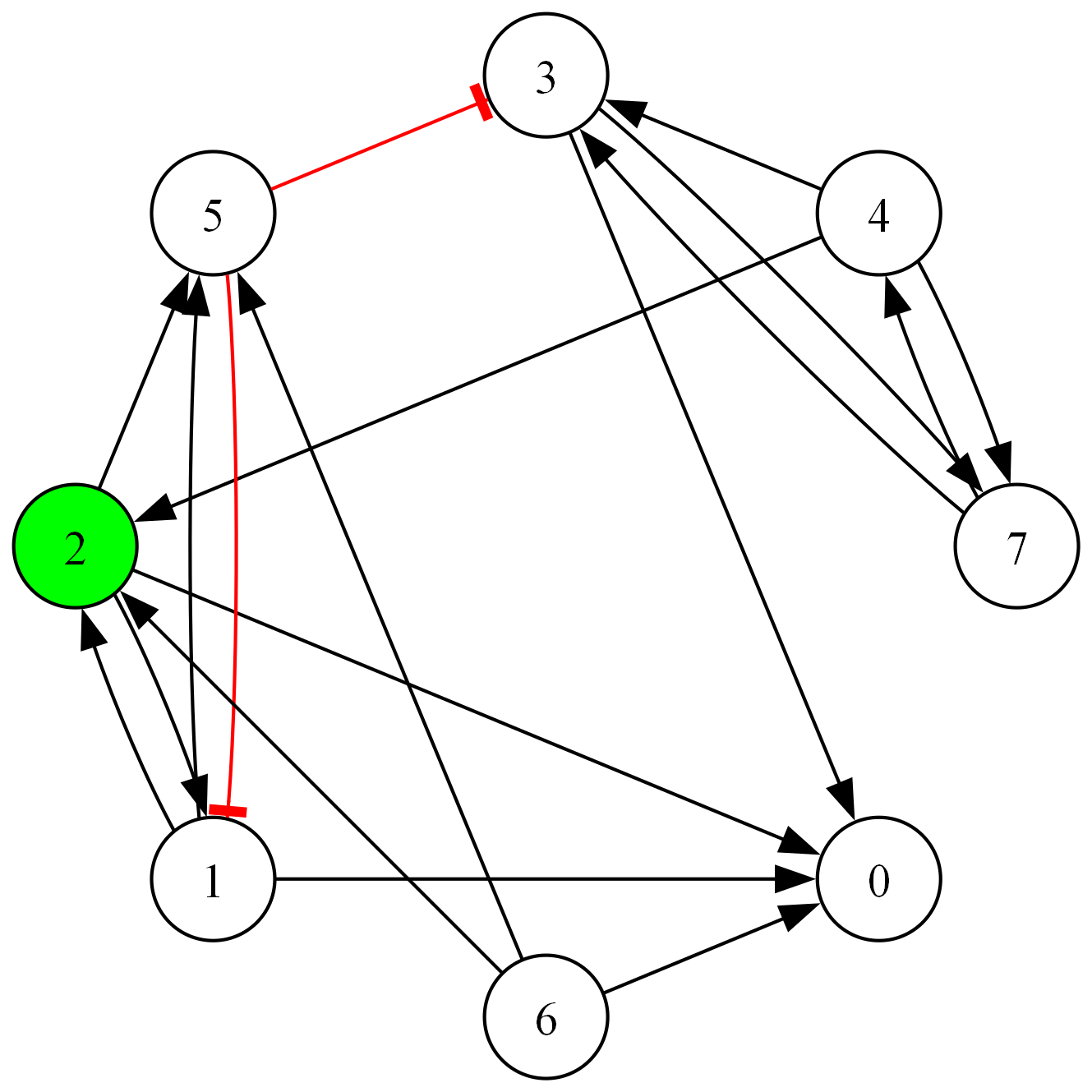}
  \caption{}
  \label{fig:circo_8cells_2}
\end{subfigure}
\begin{subfigure}{.33\textwidth}
  \centering
  \includegraphics[width=.95\linewidth]{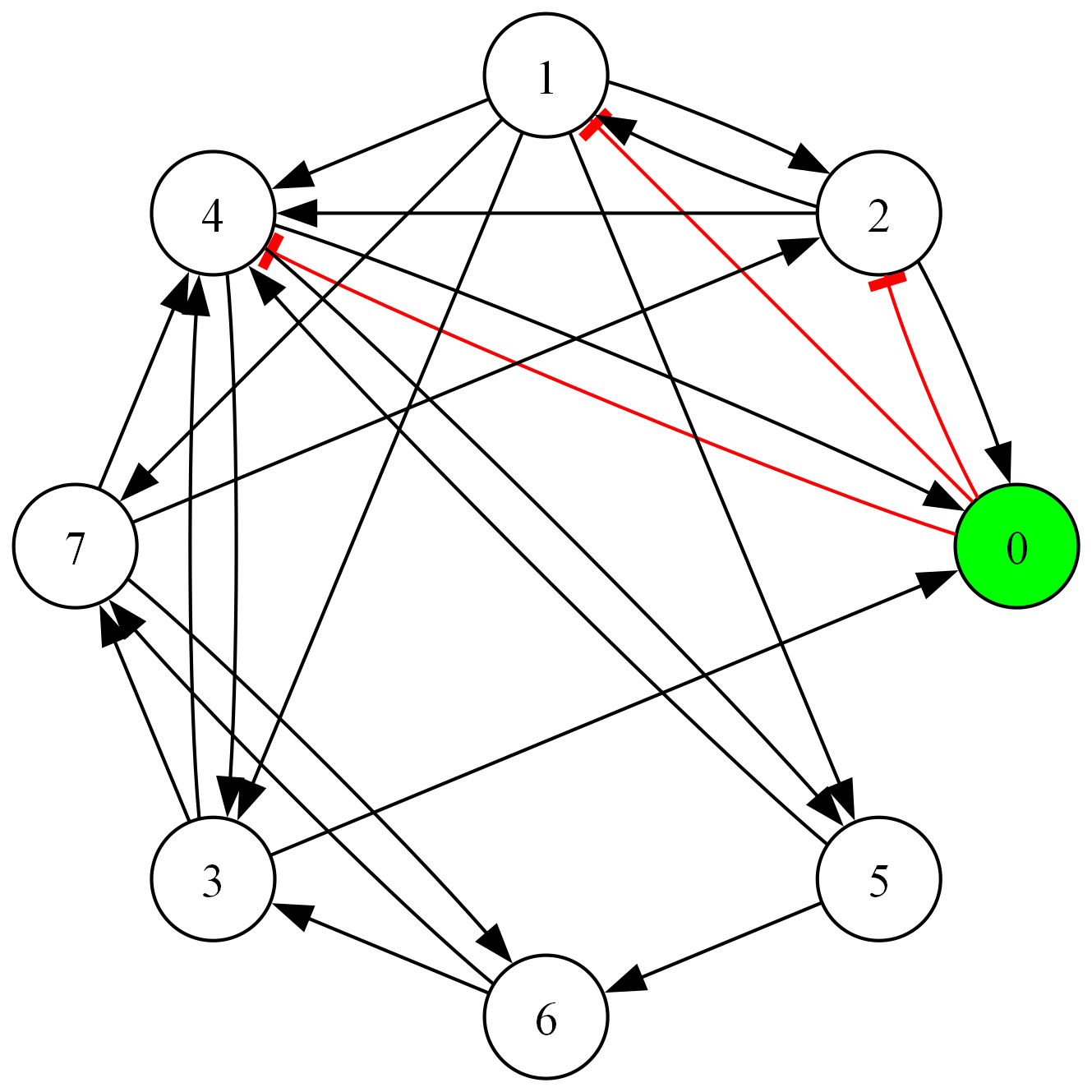}
  \caption{}
  \label{fig:circo_8cells_4}
\end{subfigure}
\begin{subfigure}{.33\textwidth}
  \centering
  \includegraphics[width=.95\linewidth]{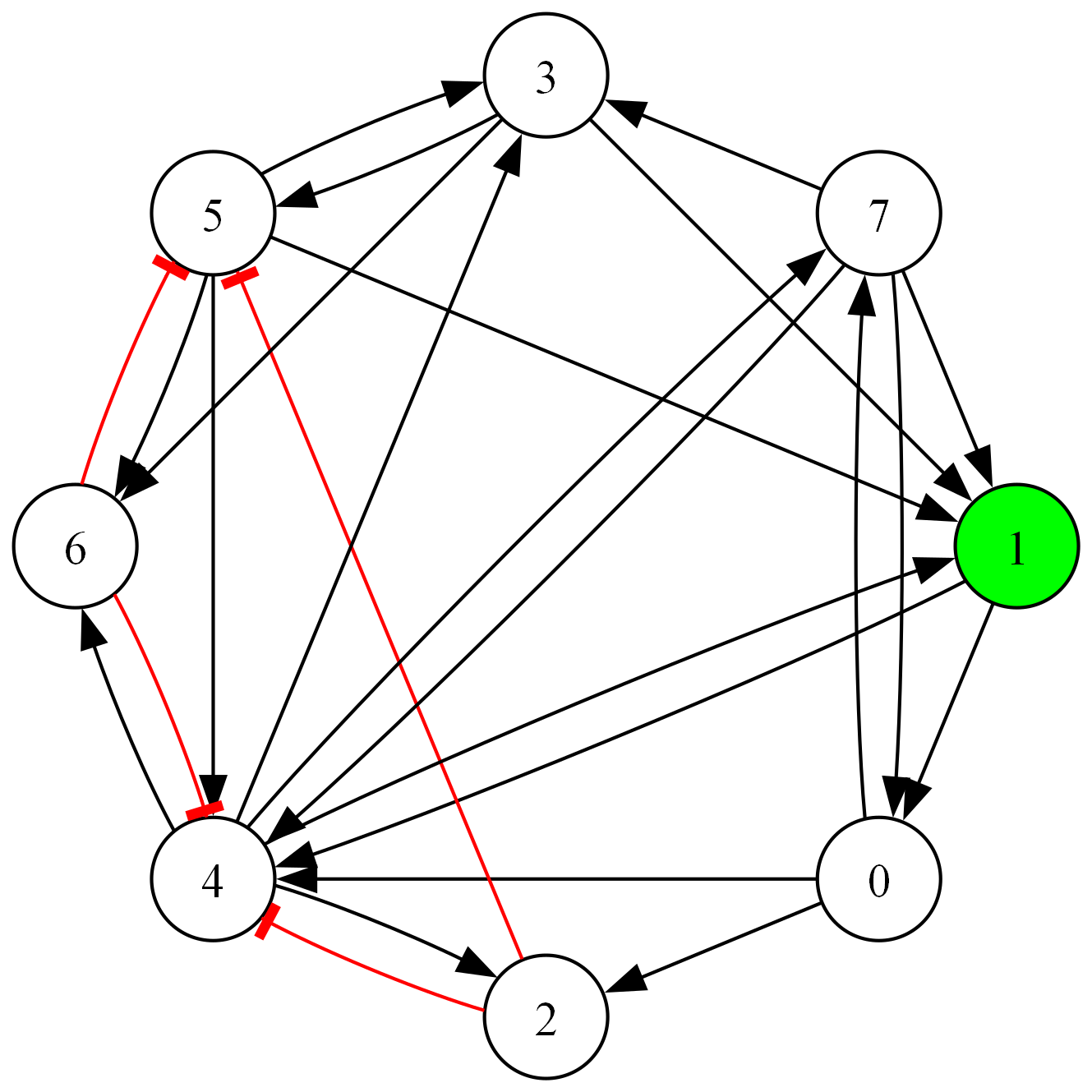}
  \caption{}
  \label{fig:circo_8cells_8}
\end{subfigure}
\caption{Three, (a), (b), and (c), instances of networks of 8 neurons that are generated as described in this section. The target neuron has a green colour. Black arrows indicate excitatory connections and red edges show inhibitory connections. A circular layout is used for visualisation purposes.}
\label{fig:circo_8cells}
\end{figure}

Network topologies, in the current work, are defined by a connectivity (i.e. adjacency) matrix that indicates for all $n$ neurons to which other neurons they are connected. Small-world networks are generated by giving neurons that are closer to each other a higher probability to form a connection, similar to experiments in \citep{SOC_reproduced}. Specifically, we make an abstraction in which neurons are placed on a line, where each pair of neurons $i, j \in \{0,...n\}$ have a probability $p_{i,j}$ to form a connection from $i$ to $j$. The distance between neuron $i$ and $j$, $d(i,j)$, is defined by $|i-j|$.
\begin{equation*}
    p_{i,j}
    \begin{cases}
      0.3 & \text{if}\ d(i,j) \leq 4, \\
      0.2 & \text{if}\ 5 \leq d(i,j) \leq 8, \\
      0.1 & \text{if}\ 9 \leq d(i,j) \leq 12, \\
      0.01 & \text{otherwise}
    \end{cases}
\end{equation*}
To maintain a ratio of 4:1 between excitatory and inhibitory neurons, each neuron has a 0.8 probability to be excitatory and 0.2 to be inhibitory. Figure~\ref{fig:circo_8cells} shows three instances of networks that are generated in this way, in a circular layout for visualisation purposes.

The target neuron is randomly assigned, with the only constraint that it has at least one incoming connection from an excitatory neuron (otherwise it is not possible to generate spikes in the target neuron by manipulating other neurons).

Due to the random generation of networks and the underlying dynamics of action potential propagation, not any spiking pattern can be created. For example, just the (short) refractory period of neurons already does not allow for a neuron to be active in every time step. Target patterns are therefore generated by simulating an episode with random actions and recording the spiking pattern that is constituted in the target neuron. In this way, we ensure that there does exist a policy that can obtain the maximum return. The only constraint for the target pattern is that there must be at least one spike present.

\section{Encoding observable context}\label{app:context_encoding}

The target neuron and target spiking pattern, $c_{\mathrm{neuron}}$ and $c_{\mathrm{pattern}}$, are observable for the agent. The parameterised Q-function can therefore additionally be conditioned on this information. A simple FF network with ReLU activation is used to encode $c_{\mathrm{neuron}}$. The target pattern, of length $N$, is encoded slightly more elaborate. The pattern is shifted in such a way that the element $(c_{\mathrm{neuron}})_t$ at current time step $t$ is in front (i.e. at the first index) when encoding the pattern with an FF network and ReLU activation. As we are constrained by a fixed input length in an FF network, the input pattern is padded with $-1$. Thus, the input of the FF network at time step $t$ is: $(c_{\mathrm{neuron}})_{t:N-1}$ concatenated with $\{-1\}^{t}$. The encoded target neuron and target pattern are thereafter concatenated to shape the encoding of the observable context.

\section{Experimental setup}\label{app:exp_setup}

The hyperparameters that are used in the experiments are listed in Table \ref{tab:rl_hyperparameters}. We used double Q-learning \citep{doubleqlearning}, target networks with soft updates, epsilon-greey exploratino and uniform sampling from the replay buffer. The hidden size in the different components of the architectures are chosen in such a way that there are no extreme differences in total number of parameters between the architectures.

\begin{table}[htbp]
    \caption{Hyperparameters used in experiments.}
    \label{tab:rl_hyperparameters}
    \centering
    \begin{tabular}{l|r|l}
        \textbf{Hyperparameter} & \textbf{Value} & \textbf{Note} \\
        \hline\hline
        Episode length & 60 & \\
        Environment steps & 1M & \\
        Replay buffer size & 16667 & All episodes are stored. \\
        Batch size & 5 & \\
        Discount factor & 0.99 & \\
        Gradient updates per batch & 10 & \\
        Target network soft update coefficient & 0.1 & \\
        Initial $\epsilon$ & 1.0 & \\
        Final $\epsilon$ & 0.05 & \\
        Anneal time & 0.25 & Fraction of total number of episodes. \\
        \hline
        Optimiser & Adam & \\
        Learning rate & $ 5 \times 10^{-4}$ & \\
        \hline
        Memory hidden size & 128 & \\
        Context hidden size & 32 & 16 target cell + 16 target pattern. \\
        Attention layers & 2 & \\
        Attention heads & 1 & \\
        Attention hidden size & 128 & \\
        Transformer FF hidden size & 256 & \\
        Non-Transformer FF hidden size & 512 & \\
        Final FF layer hidden size & [512*,256,128] & *512 not in TDRQN. \\
    \end{tabular}
\end{table}

\end{document}